\documentclass[journal]{IEEEtran}

\usepackage{url}
\usepackage{amssymb,amsmath}
\usepackage[all]{xy}
\usepackage{bm}
\usepackage[lined,boxed,commentsnumbered]{algorithm2e}
\usepackage{subfigure}
\usepackage{graphicx,color,psfrag}
\usepackage{xspace}
\usepackage{url}
\usepackage{cite,xspace,xy,soul}
\usepackage{dsfont}
\usepackage{cases}
\usepackage{psfrag}
\usepackage{pdfsync}
\usepackage{soul}
\usepackage[update,prepend]{epstopdf}
\usepackage[noend]{algorithmic}

\newcommand{\varspace}{\bm{r}, \bm{p}}

\newcommand{\sinr}{\mathsf{SINR}}

\newcommand{\sif}{\bm{I}}

\newcommand{\journal}[1]{#1}
\newcommand{\conference}[1]{}

\newcommand{\MyFigSize}{0.58}
\newcommand{\MySubFigSize}{0.22}
\newcommand{\MyFigSizetwo}{0.44}

\begin{document}

\title{On Power and Load Coupling in Cellular Networks for Energy Optimization}

\author{\IEEEauthorblockN{Chin Keong Ho}
\IEEEauthorblockA{Institute for Infocomm Research\\
 A$^*$STAR, Singapore\\
Email: hock@i2r.a-star.edu.sg}
\and
\IEEEauthorblockN{Di Yuan, Lei Lei}
\IEEEauthorblockA{Department of Science and Technology\\
Link{\"o}ping University, Sweden\\
Email:  \{di.yuan, lei.lei\}@liu.se}
\and
\IEEEauthorblockN{Sumei Sun}
\IEEEauthorblockA{Institute for Infocomm Research\\
A$^*$STAR, Singapore\\
Email: sunsm@i2r.a-star.edu.sg}}

\author{Chin Keong Ho, Di Yuan, Lei Lei, and~Sumei Sun%
\thanks{This paper is presented in part at the IEEE International Conference on Communications, June 2014.}
\thanks{C. K. Ho and S. Sun are with the Institute for Infocomm Research, A*STAR, 1 Fusionopolis Way, \#21-01 Connexis, Singapore 138632 (e-mail: \{hock, sunsm\}@i2r.a-star.edu.sg).}
\thanks{D. Yuan and L. Lei are with the Department of Science and Technology, Link{\"o}ping University, Sweden. (e-mail: \{di.yuan, lei.lei\}@liu.se)}
}

\newcommand{\Scale}{0.7}
\newcommand{\SPACING}{-1cm}

\newtheorem{conjecture}{Conjecture}
\newtheorem{remark}{Remark}
\newtheorem{insight}{Insight}
\newtheorem{question}{Question}
\newtheorem{proposition}{Proposition}
\newtheorem{corollary}{Corollary}
\newtheorem{lemma}{Lemma}
\newtheorem{assumption}{Assumption}
\newtheorem{theorem}{Theorem}
\newtheorem{example}{Example}
\newtheorem{property}[theorem]{Property}

\newcommand{\myse}{\IEEEyessubnumber} 
\newcommand{\myses}{\myse\IEEEeqnarraynumspace} 

\newcommand{\set}[1]{\mathcal{#1}}

\newcommand{\bn}{\begin{enumerate}}
\newcommand{\en}{\end{enumerate}}

\newcommand{\bi}{\begin{itemize}}
\newcommand{\ei}{\end{itemize}}

\newcommand{\be}{\begin{IEEEeqnarray}{rCl}}
\newcommand{\ee}{\end{IEEEeqnarray}}

\newcommand{\benl}{\begin{IEEEeqnarray*}}
\newcommand{\eenl}{\end{IEEEeqnarray*}}

\newcommand{\bel}{\begin{IEEEeqnarray}}
\newcommand{\eel}{\end{IEEEeqnarray}}

\newcommand{\ben}{\begin{IEEEeqnarray*}{rCl}}
\newcommand{\een}{\end{IEEEeqnarray*}}

\newcommand{\barr}{\begin{array}}
\newcommand{\earr}{\end{array}}

\newenvironment{definition}[1][Definition:]{\begin{trivlist}
\item[\hskip \labelsep {\it #1}]}{\end{trivlist}}

\newcommand{\ud}{\mathrm{d}} 

\newcommand{\FigSize}{0.6}
\newcommand{\FigSizeSmall}{0.5}

\newcommand{\avesnr} {\bar{\gamma}} 
\newcommand{\snr} {\gamma} 

\newcommand{\re}[1]{(\ref{#1})}

\newcommand{\Pe} {P_{\mathrm {e}}} 

\newcommand{\goodgap}{%
\hspace{\subfigtopskip}%
\hspace{\subfigbottomskip}}

\newcommand{\dhat}[1]{\Hat{\Hat{#1}}} 
\newcommand{\that}[1]{\Hat{\Hat{\Hat{#1}}}} 
\newcommand{\dtilde}[1]{\Tilde{\Tilde{#1}}} 
\newcommand{\ttilde}[1]{\Tilde{\Tilde{\Tilde{#1}}}} 

\newcommand{\trace}[1]{\mathrm{tr}\{#1\}} 

\newcommand{\diag}{\mathop{\mathrm{diag}}}
\newcommand{\load}{x}
\newcommand{\loadv}{\bm{x}}
\newcommand{\ymax}{Y}
\newcommand{\Pmax}{P_{\rm{max}}}
\newcommand{\pn}[1]{\bm{\bar{p}}_{#1}}

\newcommand{\bma}{\bm{a}}
\newcommand{\bmb}{\bm{b}}
\newcommand{\bmc}{\bm{c}}
\newcommand{\bmd}{\bm{d}}
\newcommand{\bme}{\bm{e}}
\newcommand{\bmf}{\bm{f}}
\newcommand{\bmg}{\bm{g}}
\newcommand{\bmh}{\bm{h}}
\newcommand{\bmi}{\bm{i}}
\newcommand{\bmj}{\bm{j}}
\newcommand{\bmk}{\bm{k}}
\newcommand{\bml}{\bm{l}}
\newcommand{\bmm}{\bm{m}}
\newcommand{\bmn}{\bm{n}}
\newcommand{\bmo}{\bm{o}}
\newcommand{\bmp}{\bm{p}}
\newcommand{\bmq}{\bm{q}}
\newcommand{\bmr}{\bm{r}}
\newcommand{\bms}{\bm{s}}
\newcommand{\bmt}{\bm{t}}
\newcommand{\bmu}{\bm{u}}
\newcommand{\bmv}{\bm{v}}
\newcommand{\bmw}{\bm{w}}
\newcommand{\bmx}{\bm{x}}
\newcommand{\bmy}{\bm{y}}
\newcommand{\bmz}{\bm{z}}

\newcommand{\bmA}{\bm{A}}
\newcommand{\bmB}{\bm{B}}
\newcommand{\bmC}{\bm{C}}
\newcommand{\bmD}{\bm{D}}
\newcommand{\bmE}{\bm{E}}
\newcommand{\bmF}{\bm{F}}
\newcommand{\bmG}{\bm{G}}
\newcommand{\bmH}{\bm{H}}
\newcommand{\bmI}{\bm{I}}
\newcommand{\bmJ}{\bm{J}}
\newcommand{\bmK}{\bm{L}}
\newcommand{\bmL}{\bm{L}}
\newcommand{\bmM}{\bm{M}}
\newcommand{\bmN}{\bm{N}}
\newcommand{\bmO}{\bm{O}}
\newcommand{\bmP}{\bm{P}}
\newcommand{\bmQ}{\bm{Q}}
\newcommand{\bmR}{\bm{R}}
\newcommand{\bmS}{\bm{S}}
\newcommand{\bmT}{\bm{T}}
\newcommand{\bmU}{\bm{U}}
\newcommand{\bmV}{\bm{V}}
\newcommand{\bmW}{\bm{W}}
\newcommand{\bmX}{\bm{X}}
\newcommand{\bmY}{\bm{Y}}
\newcommand{\bmZ}{\bm{Z}}
\maketitle

\begin{abstract}
We consider the problem of minimization of sum transmission energy in cellular networks where coupling occurs between cells due to mutual interference. The coupling relation is characterized by the signal-to-interference-and-noise-ratio (SINR) coupling model. Both cell load and transmission power, where cell load measures the average level of resource usage in the cell, interact via the coupling model. The coupling is implicitly characterized with load and power as the variables of interest using two equivalent equations, namely, non-linear load coupling equation (NLCE) and non-linear power coupling equation (NPCE), respectively.
By analyzing the NLCE and NPCE, we prove that operating at full load is optimal in minimizing sum energy, and provide an iterative power adjustment algorithm to obtain the corresponding optimal power solution with guaranteed convergence, where in each iteration a standard bisection search is employed.
To obtain the algorithmic result, we use the properties of the so-called standard interference function; the proof is non-standard because the NPCE cannot even be expressed as a closed-form expression with power as the implicit variable of interest. We present numerical results illustrating the theoretical findings for a real-life and large-scale cellular network, showing the advantage of our solution compared to the conventional solution of deploying uniform power for base stations.
\end{abstract}




%

\section{Introduction}

Data traffic is projected to grow at a compound annual growth rate of
$78\%$ from 2011 to 2016 \cite{Cisco}, fueled mainly by multimedia
mobile applications.  This growth will lead to rapidly rising energy
cost \cite{CoZeBlFeJaGoAuPe10}.  In recent years, information
communication technology (ICT) has become the fifth largest industry
in power consumption \cite{FeZi08}.  In cellular networks, in
particular, base stations consume a significant fraction of the total
end-to-end energy \cite{VeHeDePuLaJoCoMaPi11}, of which $50\%$--$80\%$
of the power consumption is due to the power amplifiers
\cite{BoCo11,GrBlFeZeImSt09}.
This observation has motivated green communication techniques for
cellular networks \cite{JoHoSu12WCL,
JoHoSu12JSAC,AdJoSuTa12IEEE,HoTanSun13,JoungHoTanSun12,JoungYuanHoSun12,Kwan12,LeiYuanHoSun13}. These
technologies include adaptive approaches such as switching off power
amplifiers to provide a tradeoff of energy efficiency and spectral
efficiency \cite{JoHoSu12WCL, JoHoSu12JSAC}, selectively turning off
base stations \cite{AdJoSuTa12IEEE}, as well as energy minimization
approaches for relay systems \cite{HoTanSun13}, OFDMA systems
\cite{JoungHoTanSun12,JoungYuanHoSun12,Kwan12}, and SC-FDMA systems
\cite{LeiYuanHoSun13}.
Extensive survey of other saving-energy approaches are highlighted in \cite{CoZeBlFeJaGoAuPe10,li2011energy,6065681}.



In this paper, we focus on the important problem of
minimizing the sum energy used for transmission in cellular networks.
Besides reducing the energy cost for transmission, minimizing the
transmission energy may lead to selection of power amplifiers with
lower power rating, hence further reducing the overhead cost involved
in turning on power amplifiers.

In a cellular network where base stations are coupled due to mutual
interference, the problem of energy minimization is challenging, as
each cell has to serve a target amount of data to its set of users, so
as to maintain an appropriate level of service experience, subject to
the presence of the coupling relation between cells. To tackle this
energy minimization problem, we employ an analytical
signal-to-interference-and-noise-ratio (SINR) model that takes into
account the load of each cell \cite{SiFuFo09,MaKo10,SiominaYuan},
where a load of a cell translates into the average level of usage of
resource (e.g., resource units in OFDMA networks) in the cell.  This
load-coupling equation system has been shown to give a good approximation for
more complicated load models that capture the dynamic nature of
arrivals and service periods of data flows in the network
\cite{FehskeFettweisICC12}, especially at high data
arrival rates. Further comparison of other approximation models concluded that the load-coupled model is accurate yet tractable \cite{fehske2013flow}.
By using this tractable model, useful insights can then be developed for the design of practical cellular systems.  In our recent works
\cite{HoYuanSun13}, we have used the load coupling
equation to maximize sum utility that is an increasing function of
the users' rates.


Previous works
\cite{SiFuFo09,MaKo10,SiominaYuan,FehskeFettweisICC12,HoYuanSun13} using the load-coupling model all assume given and fixed
transmission power. For transmission energy minimization, both power
and load become variables and they interact in the coupling model,
making the analysis more challenging. In fact, the coupling relation
between cell powers cannot be expressed in closed form even for given
cell loads.  The key aspects motivating our theoretical and
algorithmic investigations are as follows.  First, is there an insightful
characterization of the operating point in terms of load that minimizes
the sum transmission energy?  Second, given a system operating point
in load, what are the properties of the coupling system in power?
Third, even if power coupling cannot be expressed in closed form, is
there some algorithm that converges to the power solution for given
cell load?


Toward these ends, our contributions are as follows. We show that if
full load is feasible, i.e., the users' data requirements can be
satisfied, then operating at full load is optimal in minimizing sum
transmission energy (Section \ref{sec:loadopt}, Theorem~\ref{thm:optsol}).
If full load is not feasible, however, then no feasible solution
exists (Section \ref{sec:loadopt}, Corollary~\ref{cor:feasibility}).  Thus, full load is necessary and sufficient to achieve the minimum transmission energy.
\journal{Moreover, the optimal power allocation for all base stations is unique (Section \ref{sec:mainresult_powersoln},
Theorem~\ref{thm:findoptp}), and can be numerically computed based on
an iterative algorithm that can be implemented iteratively at each
base station (Section \ref{sec:optpowervect}, Algorithm~\ref{alg:power}).  To prove the algorithmic result, we make use of the properties of the so-called standard
interference function \cite{Yates95}; the proof is however non-standard, because
the function of interest does not have a closed-form
expression, and hence we use an implicit method to verify its properties.
We also characterize the load region over all
possible power allocation given some minimum target data requirements (Section \ref{sec:implementability}, Theorems~\ref{thm:cond_load}--\ref{thm:loadregionisopen}).}
Finally, we obtain numerical results to illustrate the optimality of
the full-load solution on a cellular network based on a real-life
scenario \cite{momentum}. Compared with the conventional solution
where the uniform power is used for base stations, we show the
significant advantage of the power-optimal solution in terms of meeting
user demand target and reducing the energy consumption.


The rest of the paper is organized as follows. Section~\ref{sec:model}
gives the system model of the load-coupled network.
Section~\ref{sec:prob} formulates the energy minimization problem.
Section~\ref{sec:solution} characterizes the optimality of full load, while
Section~\ref{sec:optsolncal} derives properties of the power-coupling system
and an iterative power allocation algorithm that achieves the
power solution.  Numerical results are given in Section~\ref{sec:numerical}.
Section \ref{sec:con} concludes the paper.

{\it Notations}:
We denote a column vector by a bold lower case letter, say $\bm{a}$, a matrix by a bold capital letter, say $\bm{A}$, and its $(i,j)$th element by its lower case $a_{ij}$. 
\conference{We denote $\bm{a}>\bm{0}$ and $\bm{a}>\bm{1}$ if $a_{i}>0$ and $a_{i}>1$, respectively, for all $i$; similarly for the inequality $<$.
}
\journal{We denote a {\em positive} matrix as $\bm{A}>\bm{0}$ if $a_{ij}>0$ for all $i,j$.
Similarly, we denote a {\em non-negative} matrix as $\bm{A}\geq \bm{0}$ if $a_{ij}\geq 0$ for all $i,j$.
Similar conventions apply to vectors.
Finally, $\bm{0}$ and
$\bm{1}$ denote the all-zeros and all-ones vectors of suitable
lengths.
}

\section{System Model}\label{sec:model}

\subsection{Preliminaries}

We consider a cellular network consisting of $n$ base stations that interfere with each other due to resource reuse. We focus on the downlink communication scenario where base station $i\in \mathcal{N}\triangleq\{1,\cdots, n\}$ transmits with power $p_i\geq 0$ per resource unit (in time and frequency).
We refer to cell $i$ interchangeably with base station $i$. For notational convenience, we collect all power $\{p_i\}$ as vector
$\bm{p}\geq \bm{0}$.

We assume a given association of the users to the base stations.
In this association, each base station $i$ serves one unique group of users, denoted by set $\mathcal{J}_i$, where $|\mathcal{J}_i|\geq 1.$
User $j\in \mathcal{J}_i$ is served in cell $i$ at rate $r_{ij}$ that has to be at least a rate demand of $d_{ij,\mathrm{min}}\geq 0$~nats.  Thus, $d_{ij,\mathrm{min}}$ relates to a quality-of-service (QoS) constraint.
We collect all the rates as vector $\bm{r}$ and the corresponding minimum demands as $\bm{d}_{\mathrm{min}}\geq \bm{0}$. Thus, a rate vector meets the QoS constraints if  $\bm{r}\geq \bm{d}_{\mathrm{min}}$.

\subsection{Load Coupling}\label{sec:nocomplementary}

We first consider the load coupling model for the cellular network.
We denote by $\loadv=[\load_1, \cdots,\load_n]^T$ the load in the
network, where $\bm{0}\leq\loadv\leq \bm{1}$. In LTE systems, the
load can be interpreted as the fraction of the time-frequency
resources that are scheduled to deliver data.  We model the SINR of
user $j$ in cell $i$ as \cite{SiFuFo09, MaKo10, SiominaYuan,
FehskeFettweisICC12}
\be\label{eqn:sinr}
\sinr_{ij}(\loadv,\bm{p})= \frac{p_i g_{ij}}{\sum_{k\in \mathcal{N} \backslash \{i\}} p_k g_{kj} \load_k+\sigma^2 }
\ee
where $\sigma^2$ represents the noise power and $g_{ij}$ is the
channel power gain from base station $i$ to user $j$; note that
$g_{kj}, k\neq i,$ represents the channel gain from the interfering
base stations.  The function $\mathsf{SINR}_{ij}$ depends on $x_k$ for $k\neq i$, but not on $x_i$; the dependence on the entire vector $\bm{x}$ is maintained in \eqref{eqn:sinr} for notational convenience.
The SINR model \eqref{eqn:sinr} gives a good
approximation of more complicated cellular network load models
\cite{FehskeFettweisICC12}.  Intuitively, $\load_k$ can be interpreted
as the likelihood of receiving interference from cell $k$ on all the
resource units. Thus, the combined term $(p_k g_{kj} \load_k) \in [0,
p_k g_{kj}]$ is interpreted as the average interference taken over time
and frequency for all transmissions.

Given the SINR, we can transmit reliably at the maximum rate ${\tilde r}_{ij}=B \log(1+\sinr_{ij})$~nat/s per resource block, where $B$ is the bandwidth of a resource unit and $\log$ is the natural logarithm. 
To deliver a rate of $r_{ij}$~nat for user $j$, the $i$th base station thus requires $x_{ij}\triangleq r_{ij}/{\tilde r}_{ij}$ resource units.
We assume that $M$ resource units are available.
Thus, we get the load for cell $i$ as $\load_i= \sum_{j\in\mathcal{J}_i} x_{ij}/M$, i.e.,
\be\label{eqn:xi}
\load_i 
&=&
\frac{1}{MB}
\sum_{j\in\mathcal{J}_i}\frac{r_{ij}}{\log\left(1+\sinr_{ij}(\loadv,\bm{p})\right)}
\triangleq {f}_i(\loadv)
\label{eqn:nonlinearprob0}
\ee
for $i\in \mathcal{N}$. 
Without loss of generality,
we normalize $r_{ij}$ by $MB$ in \eqref{eqn:nonlinearprob0} and so we set $MB=1$.
Let $\bm{f}(\loadv)=[f_1(\loadv), \cdots, f_n(\loadv)]^T$.
In vector form, we obtain the {\em non-linear load coupling equation} (NLCE)
\be\label{eqn:nonlinearprob}
\mathrm{NLCE}:\;\; \loadv=\bm{f}(\loadv; \varspace)
\ee
for $\bm{0}\leq\loadv\leq \bm{1}$, where we have made the dependence of the load $\loadv$ on the rate $\bm{r}$ and power $\bm{p}$ explicit.

In the NLCE, the load $\loadv$ appears in both sides of the equation and cannot be readily solved as a fixed-point solution in closed form.
Intuitively, this is because the load $x_i$ for base station $i$ affects the load $x_k$ of another base station $k\neq i$, which would then in turn affect the load $x_i$.
This difficulty in obtaining the $\loadv$ in the NLCE remains despite that the function $\mathsf{SINR}_{ij}$ (and similarly function ${f}_i$) depends on $x_k$ for $k\neq i$ but not on $x_i$.

We collect the QoS constraints as $\bm{r}\geq \bm{d}_{\mathrm{min}}$.
Without loss of generality, we assume $\bm{d}_{\mathrm{min}}$ is strictly positive, as those users with zero rate can be excluded from further consideration. Hence the power vector satisfies $\bm{p}> \bm{0}$ so as to serve all the users. Consequently, the load must be strictly positive, i.e., $\bm{0}<\loadv\leq \bm{1}$.

\section{Energy Minimization Problem}\label{sec:prob}

Our objective is to minimize the sum transmission energy given by $\sum_{i=1}^n \load_i p_i$. We note that the product $(\load_i p_i)$ measures the transmission energy used by base station $i$, because the load $\load_i$ reflects the normalized amount of resource units used (in time and frequency) while the power $p_i$ is the amount of energy used per resource unit.


The energy minimization problem is given by Problem~$P0$.
\bel{RCL} \IEEEyessubnumber\label{eqn:P0}
P0: &\;\min_{\bm{p}>0, \bm{r}>0, \bm{0}< \loadv\leq \bm{1}}\;& \loadv^T \bm{p} 
\\ \IEEEyessubnumber\label{eqn:P0c1}
&\mathrm{s.t. }& \loadv=\bm{f}(\loadv; \bm{r}, \bm{p} ) \\ \IEEEyessubnumber \label{eqn:P0c2}
&& \bm{r} \geq \bm{d}_{\mathrm{min}}.
\eel
As was mentioned earlier,  the power vector $\bm{p}$ and rate vector $\bm{r}$ vector are strictly positive to satisfy the non-trivial QoS constraint.
The load vector $\loadv$ is in fact determined by the NLCE constraint \eqref{eqn:P0c1} and thus may be treated as an implicit variable.
The second constraint \eqref{eqn:P0c2} is imposed so that the rate $\bm{r}$ satisfies the QoS constraint.

We denote an optimal solution to Problem~$P0$ as $\bm{p}^{\star}, \bm{r}^{\star}$ and the corresponding load as $\loadv^{\star}$ as determined by the NLCE.
A key challenge of Problem~$P0$ is that a positive solution pair $(\bm{p}, \bm{r})$ is considered feasible only if there exists a load such that \eqref{eqn:P0c1} holds. Whether this existence holds is not obvious due to the non-linearity of the NLCE.
As such, the convexity of the optimization problem cannot be readily established, and hence standard convex optimization techniques do not apply readily. 

\section{Optimality of Full Load}\label{sec:solution}

\journal{In Section~\ref{sec:Satisfiability} and Section~\ref{sec:Implementability}, we consider fundamental properties of rate and load, respectively, such that there exists a power satisfying the NLCE.  To study the fundamental properties, we consider the existence of a load satisfying $\loadv>\bm{0}$. The additional constraint that $\loadv\leq\bm{1}$ is taken into account in Section~\ref{sec:loadopt}, in which we prove the key result that full load, i.e., $\loadv=\bm{1}$, is a necessary and sufficient condition for the solution in Problem~$P0$ to be optimal.}

\conference{In this section, we shall show that sum energy is minimized if the cellular network is at full load.}

\subsection{Satisfiability of Rate}\label{sec:Satisfiability}

We first establish conditions on rate vector $\bm{r}$ such that a load $\loadv>\bm{0}$ exists and satisfies the NLCE, possibly
with $\loadv>\bm{1}$.  We denote the spectral radius of matrix
$\bm{A}$ as $\rho(\bm{A})$, defined as the absolute value of the
largest eigenvalue of $\bm{A}$.
\begin{lemma}\label{lem:uniqueness}
For any power $\bm{p}>0$, there exists a unique load $\loadv>\bm{0}$ satisfying the NLCE if and only if
\be\label{eqn:spectralradius}
\rho(\bm{\Lambda}(\bm{r})) < 1
\ee
where the $(i,k)$th element of $\bm{\Lambda}(\bm{r})$ is given by
\be
\lambda_{ik}=
\left \{
\begin{array}{ll}
0, & \mbox{if } i=k; \\
\sum_{j\in\mathcal{J}_i}  {g_{kj} r_{ij}}/{g_{ij}}, & \mbox{if } i\neq k
\end{array}
\right .
\ee
which is a function of $\bm{r}$ (but not $\bm{p}$).
\end{lemma}
\begin{IEEEproof}
Follows directly from  \cite[Theorem~1]{HoYuanSun13}.
\end{IEEEproof}

Due to Lemma~\ref{lem:uniqueness}, we say that the rate vector
$\bm{r}$ is {\em satisfiable} if $\rho(\bm{\Lambda}(\bm{r})) < 1$. If $\bm{r}$ is not satisfiable, then there does not exist any
power $\bm{p}>\bm{0}$ that results in a load satisfying constraint
\eqref{eqn:P0c1}.  We note that even if $\bm{r}$ is satisfiable, it is still
possible that the load does not satisfy $\loadv \leq \bm{1}$ and hence
violates its upper bound.  Thus, satisfiability is a necessary
condition for a feasible solution to exist in Problem $P0$, but
it may not be sufficient.

Henceforth, we assume that a rate $\bm{r}$ is satisfiable; otherwise no feasible solution exists in Problem $P0$.
Given $\bm{p}$, we can then numerically obtain $\loadv$ by the {\em iterative algorithm for load} (IAL) \cite[Lemma~1]{HoYuanSun13}, as follows. Specifically, starting from an arbitrary initial load $\loadv^0> 0$, define the output of the $\ell$th algorithm iteration as
\be\label{eqn:algo}
\loadv^{\ell}=\bm{f}(\loadv^{\ell-1}; \varspace)
\ee
for $\ell=1,2,\cdots,L$, where $L$ is the total number of iterations.
Then $\loadv^L$ converges to the fixed-point solution $\loadv$ of the NLCE as $L$ goes to infinity.
The IAL is derived using \cite{Yates95} by showing that $\bm{f}$ is a so-called standard interference function, to be defined in Section~\ref{sec:sif}.

\subsection{Implementability of Load}\label{sec:Implementability}

Although Lemma~\ref{lem:uniqueness} states that any given power vector
$\bm{p}$ always corresponds to a load vector $\loadv$ that satisfies
the NLCE, the reverse is not true.
\journal{To obtain some intuition why this inverse mapping may fail, let us consider the special case of $n=2$ base stations with channel gain $g_{ij}=1$ for all $i,j$, rate $\bm{r}=\bm{1}$, and noise variance $\sigma^2=1$. We randomly choose the power $\bm{p}=[p_1, p_2]^T$ using a uniform distribution over $0< p_i\leq 2, i=1,2,$ which is plotted in Fig.~\ref{fig:1a}. The corresponding load $\loadv=[\load_1, \load_2]^T$ obtained using the IAL is shown in Fig.~\ref{fig:1b}.  We see that indeed there is a load region that does not appear to correspond to any power $\bm{p}> \bm{0}$.


\begin{figure}
\centering
\subfigure[][Power region.]
{\includegraphics [scale=\MySubFigSize] {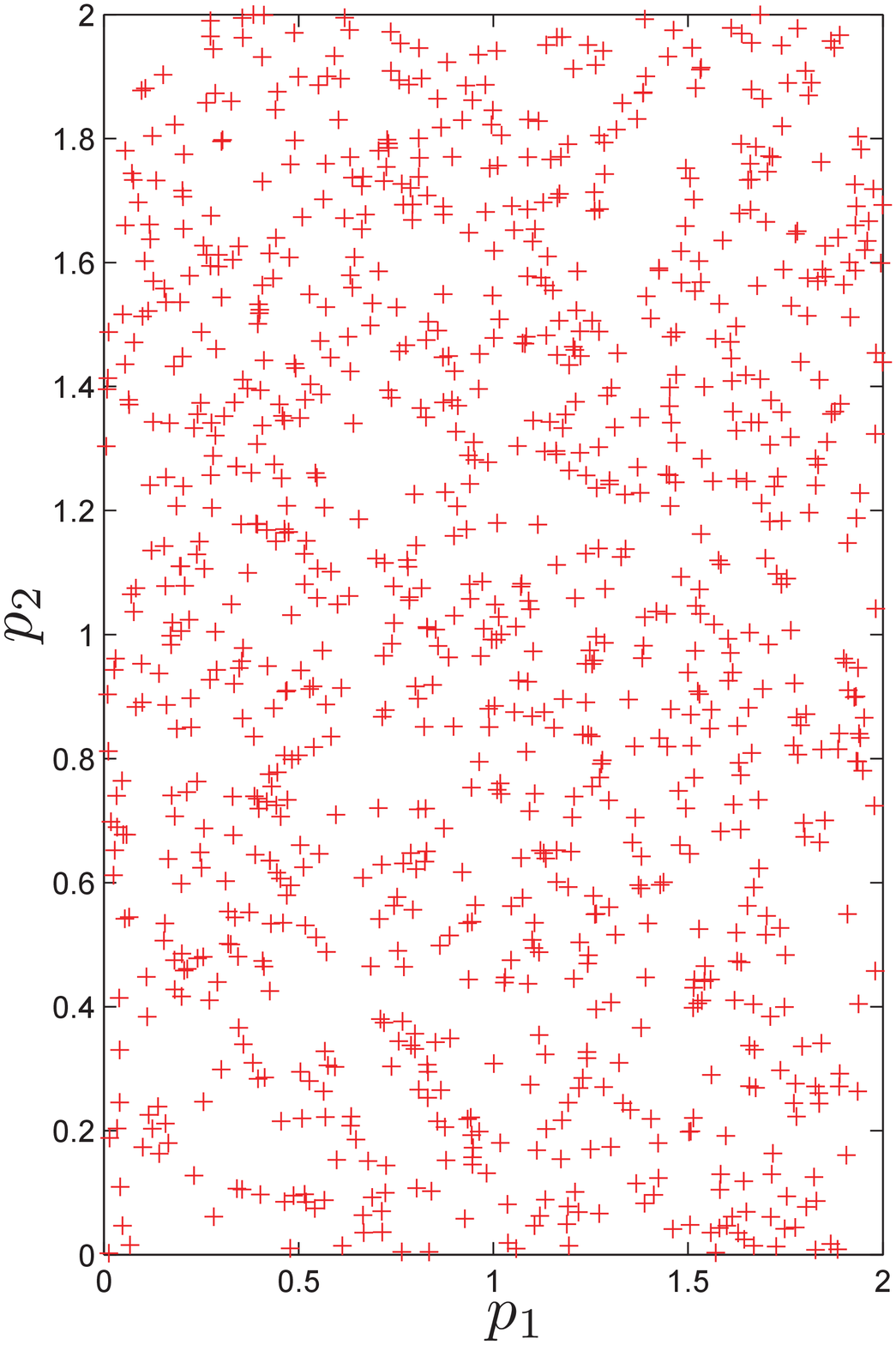}                \label{fig:1a}}
\subfigure[][Load region.]
{\includegraphics [scale=\MySubFigSize] {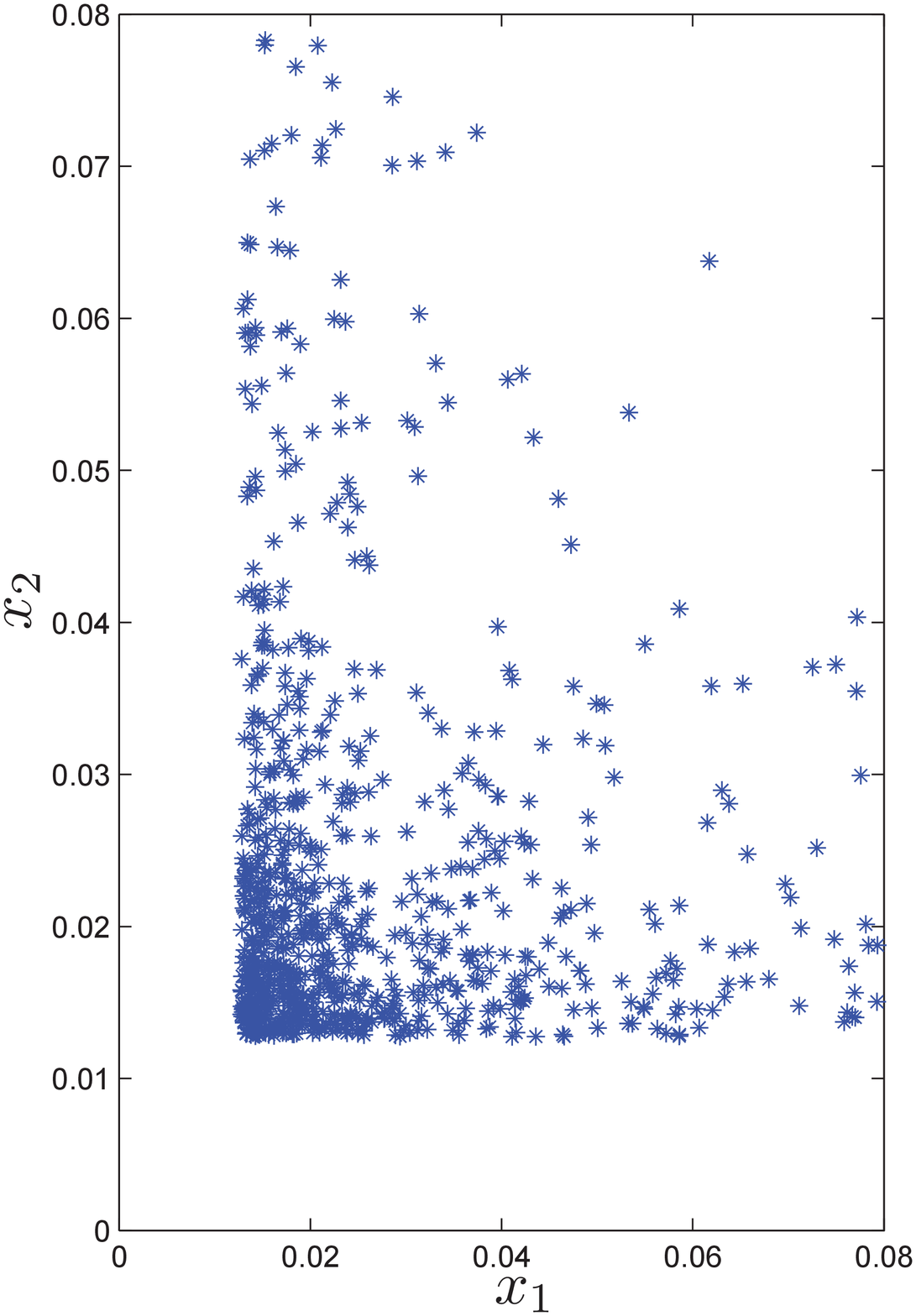}
\label{fig:1b}}
\caption{Corresponding power region and load region satisfying the NLCE.}\label{fig:1}
\end{figure}


}

Given that $\bm{r}$ is satisfiable, we say that a load $\loadv$ is
{\em implementable} if there exists power $\bm{p}$ such that the NLCE
is satisfied.

The following toy scenario shows that full load may not always be implementable.
In practice, this may occur during peak times in cellular hot spots, such as train stations, when mobile data cannot be sustained at high speeds even if all time-spectrum resources are used no matter how power is set (cf. Lemma~ \ref{lem:uniqueness}).

We assume $n=2$ cells with one user per cell, each with rate $r=2$~nat, and $\loadv=\bm{1}$. The channel gains from a base station to the user it serves and the user it does not serve is set as $g=1$ and $g'=1/3$, respectively.
Note that the rate is satisfiable since we get
$\bm{\Lambda}=\left[
\begin{array}{cc}
  0 & g'r/g \\
  g'r/g & 0
\end{array}\right]$ and hence $\rho(\bm{\Lambda})=4/9$ which satisfies \eqref{eqn:spectralradius}.
By symmetry, the power allocated for all cells must be the same
with $p_1=p_2=p$ and must thus satisfy \eqref{eqn:xi}, i.e.,
$\log(1+g p/(g' p +\sigma^2))=r$.
For any $p\geq 0$, the left hand side is upper bounded by $\log(1+gp/(g'p+\sigma^2))\leq \log(1+g/g')=\ln(4)=1.39$~nat, which is less than $r=2$~nat.
Hence, regardless of the power allocation, \eqref{eqn:xi} cannot hold and so full load is not implementable in this case.

\subsection{Main Result: Full Load is Optimal}
\label{sec:loadopt}

Our first main result is given by Theorem~\ref{thm:optsol}, which states that full load, if implementable, is optimal to minimize the sum energy in Problem~$P0$.

Lemma~\ref{thm:fullload_is_opt} given below is a key step to prove Theorem~\ref{thm:optsol}.
\begin{lemma}\label{thm:fullload_is_opt}
For Problem~$P0$, the optimal solution is such that the load vector satisfies $\loadv^{\star}=\bm{1}$. 
\end{lemma}
\begin{IEEEproof}
Note that the load satisfies $\load_i>0$ for all cell $i$ to satisfy non-trivial rate demands.
Assume that at optimality, we have $\bm{0}< \loadv^{\star}\leq \bm{1}$ where there exists at least one cell $i\in\mathcal{N}$ with load $0< x^{\star}_i< 1$ and power $p^{\star}_i$. With all other power $p^{\star}_k$ and load $\load^{\star}_k$ fixed, $k\neq i$, we reduce the power $p^{\star}_i$ to $p'=p^{\star}_i-\epsilon, \epsilon>0$. Using \eqref{eqn:nonlinearprob0}, the corresponding load $\load^{\star}_i$ strictly increases to $x'=x^{\star}_i+\epsilon', \epsilon'>0$. We choose $\epsilon>0$ such that $\load'\leq 1$. With this new power-load pair $(p',\load')$ for cell $i$, we claim that (see proofs below): (i) the objective function is reduced, and (ii) the corresponding rate vector $\bm{r}'$ is such that $\bm{r}'\geq \bm{r}^{\star}$, i.e., the NLCE constraint is satisfied since $\bm{r}^{\star} \geq \bm{d}_{\mathrm{min}}$. The two claims together imply that $\loadv^{\star}$ with $0< x^{\star}_i< 1$ is not optimal, independent of the actual cell $i$. By contradiction, $x^{\star}_i= 1$  for all $i$, i.e, $\loadv^{\star}=\bm{1}$.

We now prove the first claim. Denote the energy used in cell $i$, as a function of its power $p_i$, as
$e_i=\load_i p_i
= \sum_{j\in\mathcal{J}_i}\frac{r_{ij} p_i}{\log\left(1+c_{ij} p_i\right)}
$
where $c_{ij}\triangleq g_{ij} /(\sum_{k\in \mathcal{N} \backslash \{i\}} p_k g_{kj} \load_k+\sigma^2)$ does not depend on $p_i$ nor $x_i$.
Then
\be\label{eqn:dec_energy}
\frac{\partial e_i}{\partial p_i}&=&\sum_{j\in\mathcal{J}_i}r_{ij} \frac{(1+c_{ij}p_i) \log(1+c_{ij}p_i)-c_{ij}p_i}{\log^2(1+c_{ij} p_i)(1+c_{ij}p_i)}.
\ee
It can be verified by calculus that the numerator of each summand is strictly increasing for $p_i\geq 0$. Since the numerator equals zero at $p_i=0$, the numerator is strictly positive for $p_i>0$. Clearly the denominator is strictly positive for $p_i>0$. Thus, $\frac{\partial e_i}{\partial p_i}>0$.
Hence, when the power for cell $i$ is decreased, the energy $e_i$ decreases. Thus, the objective function also decreases.

To prove the second claim, we first note that for cell $i$, we have constrained the new power-load pair $(p',\load')$ to satisfy \eqref{eqn:nonlinearprob0}. Thus, the new rate for cell $i$, denoted by $r_{ij}', j\in \mathcal{J}_i$, is the same as the optimal rate $r_{ij}^{\star}$ corresponding to the power-load pair $(p^{\star}_i,\load^{\star}_i)$.
Next, we observe that the product $\load'p'$ is strictly smaller as compared to $\load^{\star}_i p^{\star}_i $, according to the first claim. Thus, for user $j\in \mathcal{J}_k$ in cell $k\neq i$, $\sinr_{kj}(\loadv)$ strictly increases. It follows that the NLCE for cell $k$ is satisfied with the same load $\load_k$ but with a larger rate $r_{kj}'$ as compared to the optimal rate $r_{kj}^{\star}$. In summary, we thus have $\bm{r}'\geq \bm{r}^{\star}$.
\end{IEEEproof}

\begin{theorem}\label{thm:optsol}
Suppose full load, i.e., $\loadv=\bm{1}$, is implementable. 
Then the optimal solution for Problem~$P0$ is as follows: $\bm{r}^{\star} = \bm{d}_{\mathrm{min}}$, and $\bm{p}^{\star}$  is such that $\loadv^{\star}=\bm{1}$. The optimal power vector $\bm{p}^{\star}$ is thus given implicitly by the NLCE as
\be\label{eqn:optpow}
\bm{1}=\bm{f}(\bm{1}; \bm{d}_{\mathrm{min}}, \bm{p}^{\star}).
\ee
\end{theorem}
\begin{IEEEproof}
The proof follows from Lemma~\ref{thm:fullload_is_opt} and Lemma~\ref{thm:constraintequality} given in the Appendix, which state that $\loadv^{\star}=\bm{1}$ and $\bm{r}^{\star} = \bm{d}_{\mathrm{min}}$ are the optimal solutions, respectively.
Substituting the optimal solutions into the NLCE results in \eqref{eqn:optpow}.
\end{IEEEproof}

From Theorem~\ref{thm:optsol}, serving the minimum required rate is
optimal. This observation is intuitively reasonable as less
resources are used and hence less energy is consumed. Interestingly,
Theorem~\ref{thm:optsol} states that having full load is optimal. This
second observation is not as intuitive, since it is not immediately
clear the effect of using higher load on both the sum energy and
interference. This is because using a high load may lead to more
interference to neighbouring cells, which may then require other cells
to use more energy to serve their users' rates. Mathematically, the
reason can be attributed to the proof of
Lemma~\ref{thm:fullload_is_opt}, which shows that, as the power
decreases, the energy as well as the interference for each cell
decreases, while concurrently the load increases. Thus, by using full
load, the energy is minimized.

\journal{
Next, Corollary~\ref{cor:feasibility} provides a converse type of result to Theorem~\ref{thm:optsol}.
The result follows from a theorem with a generalized statement, which we defer to Section~\ref{sec:mainresult_powersoln} because the proof requires the use of algorithmic notions for finding power given load.
\begin{corollary}\label{cor:feasibility}
If full load $\loadv=\bm{1}$ is not implementable, then there is no other load ${\loadv}\leq \bm{1}$ with ${\loadv}\neq \bm{1}$ that is implementable. Thus, there is no feasible solution for Problem~$P0$.
\end{corollary}
\begin{IEEEproof}
The result follows as a special case of Theorem~\ref{thm:cond_load} later in Section~\ref{sec:mainresult_powersoln}.
\end{IEEEproof}

\begin{remark}
Theorem~\ref{thm:optsol} and Corollary~\ref{cor:feasibility} together thus show that full load is both necessary and sufficient to achieve the minimum energy in Problem~$P0$.
\end{remark}
}


\journal
{\begin{remark}\label{rem:generealobjfn}
It can be easily checked that Theorem~\ref{thm:optsol} and Corollary~\ref{cor:feasibility} continue to hold even if we generalize the objective function to any function $c(\load_1 p_1, \load_2 p_2, \cdots, \load_n p_n)$ that is increasing in each of its argument. For example, $c(y_1,\cdots,y_n)=\sum w_i y_i$ gives the weighted sum energy with positive weights $\{w_i, i =1, \cdots, n\}$.
\end{remark}
}

\journal{
\section{Optimal Power Solution}\label{sec:optsolncal}

Although full load is optimal for Problem~$P0$, it is still not clear
if the optimal power $\bm{p}^{\star}$ is unique and how to numerically
compute $\bm{p}^{\star}$ in
\eqref{eqn:optpow}. Our second main result, Theorem~\ref{thm:findoptp}, answers both
questions, but in a more general setting. Namely, we provide
theoretical and algorithmic results for finding power $\bm{p}$ given
arbitrary load $\loadv$ that is implementable (not necessarily all ones) and arbitrary rate $\bm{r}$ that is satisfiable (not necessarily equal to
$\bm{d}_{\mathrm{min}}$), so as to satisfy the NLCE.

\subsection{Standard Interference Function}\label{sec:sif} 

Before we state the main result of the section, we recap the standard interference function and the iterative algorithm introduced in \cite{Yates95}. The algorithm shall be used to obtain the optimal power $\bm{p}^{\star}$, and is also a key step in the proof of the implementability of load.

Consider a function $\sif: \mathbb{R}^n_+ \rightarrow \mathbb{R}^n_+$.
We denote the input as $\bm{p}$ as we shall focus on using power as the input.
We say $\sif(\cdot)$ is a standard interference function if it satisfies the following properties for all input power $\bm{p}\geq \bm{0}$ \cite{Yates95}.
\bn
\item Positivity: $\sif(\bm{p})>\bm{0}$;
\item Monotonicity: If $\bm{p}\geq \bm{p}'$, then $\sif(\bm{p})\geq \sif(\bm{p}')$.
\item Scalability: For all $\alpha>1$, $\alpha \sif(\bm{p}) > \sif(\alpha\bm{p})$.
\en

Next, we consider both the synchronous and asynchronous versions of the
{\em iterative algorithm for power} (IAP), similar to the two versions of iterative algorithm in \cite{Yates95}. IAP generates a sequence of power vectors via multiple iterations. In each iteration, the power vector produced amounts to evaluating function $\sif(\cdot)$ with the previous iterate as the input. As power is a vector, when the calculation of one power element is performed, there is a choice of whether or not to use this updated power value in the function evaluation for the remaining power elements. These two choices lead to the synchronous and asynchronous IAPs. We consider a specific form of asynchronous IAPs which will turn out to be useful for our proof of Theorem~\ref{thm:cond_load} later.

We assume $L$ iterations are performed in each case.
For the {\em synchronous} IAP, the entire power vector is updated in each iteration. In contrast, for the {\em asynchronous} IAP, there are $n$ inner iterations for each (outer) iteration, and in each inner iteration, only one power element is updated.

\bi

\item  Synchronous IAP: Assume an arbitrary initial power given by $\bm{p}^0> \bm{0}$.
The output for iteration $\ell=1,\cdots,L$, is given by
\be\label{eqn:IAP}
\bm{p}^{\ell}=\sif(\bm{p}^{\ell-1}).
\ee
Clearly, any power element of $\bm{p}^{\ell}$ is solely determined by
$\bm{p}^{\ell-1}$.

\item  Asynchronous IAP:
In each iteration $\ell=1,\cdots,L$,  we perform $n$ inner iterations.
Assume an arbitrary initial power given by $\bm{p}^0_0> \bm{0}$.
The output of the
$i$th inner iteration, $i=1,\cdots, n,$ is given by
\be\label{eqn:IAPasy}
p_i^{\ell}=\sif(\bm{p}_{i-1}^{\ell-1})
\ee
where $\bm{p}_{i-1}^{\ell-1}$ represents the power vector containing the most current elements after $(\ell-1)$ outer iterations and $(i-1)$ inner iterations (i.e., during the $\ell$th iteration).
After $L$ (outer) iterations are fully completed, each with $n$ inner iterations, we obtain $\bm{p}_{n}^{L}$ as the final power vector solution.
\ei

Lemma~\ref{lem:yates}  demonstrates the use of the IAP algorithms to obtain the unique fixed-point point solution.
\begin{lemma}\label{lem:yates}
Suppose a fixed-point solution $\bm{p}$ exists for $\bm{p}=\sif(\bm{p})$.
If $\sif$ is a standard interference function, then starting from any initial power vector, both the synchronous and asynchronous IAP algorithms converge to the fixed-point solution $\bm{p}$, which is unique.
\end{lemma}
\begin{IEEEproof}
We omit the proof which is found in \cite[Theorems~2,4]{Yates95}.
\end{IEEEproof}

\subsection{Main Result: Existence and Computation of Power Solution}
\label{sec:mainresult_powersoln}

Before proving the main result, we present and prove some properties
on how the elements of the power vector relate to each other in
NLCE. The properties will then be used to establish that the results in
\cite{Yates95} with the notion of standard interference
function can be applied.

Let $\pn{i}$ be the vector of length $(n-1)$ that contains all elements in
vector $\bm{p}$ except for element $p_i$. For example, if $\bm{p}=[p_1, p_2, p_3,\cdots, p_n]$, then  $\pn{2}=[p_1, p_3, \cdots, p_n]$.
Lemma~\ref{lem:property1} shows that given
$\loadv$ and $\bm{r}$, the dependency of $p_i$ on $\pn{i}$ (such that the NLCE holds) qualifies as a function,
even if the function is not in closed form.

\begin{lemma}\label{lem:property1}
Let $\bm{p}, \loadv, \bm{r}$ satisfy the NLCE, where the vectors are strictly positive.
Then there exists function $h_i:\mathbb{R}^n_{++}\rightarrow \mathbb{R}^n_{++}$ satisfying $p_i=h_i(\pn{i}; \loadv, \bm{r})$ for all $i=1,\cdots, n$. Writing $p_i$'s and $h_i$'s in vector form, we get $\bm{p}=\bm{h}(\bm{p}; \loadv, \bm{r})$.
\end{lemma}
\begin{IEEEproof}
We fix $\loadv, \bm{r}$ and drop these notations in the function $h_i(\cdot)$ for simplicity.
To prove the existence of the function $h_i(\cdot)$, we need to show that given $\pn{i}$, there exists a unique $p_i$ for $i=1,\cdots, n$. First, we write the NLCE in \eqref{eqn:nonlinearprob0}  as
\be\label{eqn:nlce1}
1=\sum_{j\in\mathcal{J}_i}\frac{a_{ij}}{\log\left(1+p_i b_{ij}(\pn{i}, \sigma^2) \right)} \triangleq \eta_i(p_i)
\ee
where
\be\label{eqn:a}
a_{ij}&\triangleq& r_{ij}/\load_{i} \\
\label{eqn:b}
b_{ij}(\pn{i}, \sigma^2)&\triangleq & \frac{g_{ij}}{\sum_{k\in \mathcal{N} \backslash \{i\}} p_k g_{kj} \load_k+\sigma^2 }
\ee
are both independent of $p_i$.
We fix $\pn{i} >0$ and $\sigma^2\geq 0$. It follows that $b_{ij}(\pn{i}, \sigma^2)>0$ and so $\eta_i(p_i)>0$.
Observe that $\eta_i(p_i)$ is a strictly decreasing function of $p_i$. Since $\eta_i(p_i)\rightarrow \infty$ as $p_i\rightarrow 0$, and  $\eta_i(p_i)\rightarrow 0$ as $p_i\rightarrow \infty$, there  exists a unique $p_i>0$ such that $\eta_i(p_i)=1$, and thus satisfies \eqref{eqn:nlce1}.
Hence there exists a function of the form  $p_i=h_i(\pn{i})$, for any $i$.
\end{IEEEproof}

\begin{remark}\label{rem:noclosedform}
The function $h_i(\cdot)$ does not submit to a closed-form solution. For example, consider expressing $p_i$ in terms of $\pn{i}$ in  \eqref{eqn:nlce1} where the number of summands is $|\mathcal{J}_i|>1$. Because each of them is non-linear in $p_i$, the dependency of $p_i$ on $\pn{i}$ is not explicit.
\end{remark}

\begin{remark}\label{rem:computeh}
Although $h_i(\cdot)$ cannot be expressed in closed form, we can
numerically obtain the output $p_i$ of the function $h_i$ given the
input $\pn{i}$. Equivalently, this means that we want to obtain the
value of $p_i$ such that \eqref{eqn:nlce1} holds. This is computed,
for example, by a bisection search on $\eta_i(p_i)=1$, making use of the
property that $\eta_i(p_i)$ is a strictly decreasing function.
Specifically, we first choose an arbitrary but small power $p'$ such
that $\eta_i(p')>1$ and an arbitrary but large power $p''$ such that
$\eta_i(p'')<1$. Next we use the new power $p=(p'+p'')/2$ and evaluate if
$\eta_i(p)$ is greater or smaller than one, then replace $p'$ or $p''$ by
$p$, respectively. By performing this procedure iteratively, we have
guaranteed convergence to the desired $p$ that satisfies $\eta_i(p_i)=1$.
This forms the basis for the proposed algorithm later in
Section~\ref{sec:numerical}.
\end{remark}

We observe that $\bm{h}(\cdot)$ is to some extent similar to
$\bm{f}(\cdot)$ in the NLCE \eqref{eqn:nonlinearprob}.  From
Remark~\ref{rem:noclosedform}, however, the function $\bm{h}(\cdot)$ cannot be
readily written as a closed-form expression.  Thus, proving properties
related to $\bm{h}(\cdot)$ is more challenging, as compared to the
case of $\bm{f}(\cdot)$ for which a closed-form solution is available.
Nevertheless, Lemma~\ref{lem:property2} states that $\bm{h}(\cdot)$
qualifies as a standard interference function as defined in
\cite{Yates95}.

\begin{lemma}\label{lem:property2}
Given load $\loadv\geq \bm{0}$ and rate $\bm{r}\geq \bm{0}$, $\bm{h}(\bm{p};\loadv, \bm{r})$ is a standard interference function in $\bm{p}$.
\end{lemma}
\begin{IEEEproof}
Henceforth we assume that load $\loadv\geq \bm{0}$ and rate $\bm{r}\geq \bm{0}$ are given.
For notational convenience, we drop the dependence of these entities in the notation of $\bm{h}(\cdot)$.
We consider an arbitrary $i$ and refer to $\eta_i(p_i), a_{ij}, b_{ij}$ as defined in \eqref{eqn:nlce1}, \eqref{eqn:a} and \eqref{eqn:b}, respectively, throughout the proof. For this proof, it is useful to denote the function $\eta_i(p_i)$ explicitly as $\eta_i(p_i,\pn{i}, \sigma^2)$ to ease the discussion.
We prove each of the three properties required for standard interference function below.

{\em Positivity}: From the proof of Lemma~\ref{lem:property1}, there exists a unique $p_i>0$ that satisfies \eqref{eqn:nlce1}, i.e., $h_i(\pn{i})>0$. This holds for all $i$, thus $\bm{h}(\bm{p})>\bm{0}$.

{\em Monotonicity}: From \eqref{eqn:nlce1}, we observe that
$\eta_i(p_i,\pn{i},\sigma^2)$ strictly increases as $p_i$ decreases, or
as any element of $\pn{i}$ increases. Hence, to satisfy
$\eta_i(p_i,\pn{i},\sigma^2)=1$, $p_i$ strictly increases if any
element of $\pn{i}$ increases. We note that an equivalent
representation of $\eta_i(p_i,\pn{i},\sigma^2)=1$ is
$p_i=h_i(\pn{i})$. It follows that $h_i(\pn{i})$ is increasing in any
of the arguments.

{\em Scalability}:
Let $q_1=h_i(\pn{i})$ and  $q_2=h_i(\alpha\pn{i})$, where $\alpha>1$. Observe that
\be\label{eqn:temp1}
    \eta_i(q_1,\pn{i},\sigma^2)=\eta_i(q_2,\alpha \pn{i}, \sigma^2)
\ee
since both equal one according to \eqref{eqn:nlce1}.
It is easy to check that $q_1 b_{ij}(\pn{i}, \sigma^2)=\alpha q_1 b_{ij}(\alpha\pn{i}, \alpha\sigma^2)$.
That is, multiplying all the terms in the triplet $(q_1,\pn{i},\sigma^2)$ by a positive constant still allows
 \eqref{eqn:nlce1} to be satisfied.
Thus we get from \eqref{eqn:temp1}
\be\label{eqn:temp2}
    \eta_i(\alpha q_1,\alpha\pn{i},\alpha\sigma^2)=\eta_i(q_2,\alpha \pn{i}, \sigma^2).
\ee
With the second argument in $\eta_i(y,\cdot, z)$ fixed, we note that the
output of the function strictly decreases with $y$ and strictly
increases with $z$.  By the equality in \eqref{eqn:temp2}, it
follows that $\alpha q_1> q_2$ because
$\alpha\sigma^2>\sigma^2$. Taking into account of the definition of
$q_1$ and $q_2$, we have proved $\alpha
h_i(\pn{i})>h_i(\alpha\pn{i})$.
\end{IEEEproof}


Using Lemma~\ref{lem:property1} and Lemma~\ref{lem:property2}, we are
ready to provide the main result, stating that NLCE can be expressed
in an alternative form with the power taken as the subject of
interest.  The proof is non-standard, because the relations among the
power elements do not submit to a closed form
(Remark~\ref{rem:noclosedform}). Hence, it has been necessary to first
establish that the relation between one power element and the others
qualifies as a function (Lemma~\ref{lem:property1}). Next,
we have used an implicit method to prove that
$\bm{h}(\cdot)$ is indeed a standard interference function
(Lemma~\ref{lem:property2}).

\begin{theorem}\label{thm:findoptp}
Given load $\loadv$ and rate $\bm{r}$, the power $\bm{p}$ that satisfies the NLCE can be represented equivalently
in the form of a non-linear {\em power} coupling equation (NPCE) given by%
\be\label{eqn:NPCE}
\mathrm{NPCE}:\;\; \bm{p}=\bm{h}(\bm{p}; \loadv, \bm{r})
\ee
where $\bm{h}(\cdot)$ is a standard interference function. Given that a solution $\bm{p}$ exists, then $\bm{p}$ is unique and can be obtained numerically by the IAP.
\end{theorem}
\begin{IEEEproof} 
Lemma~\ref{lem:property1} states the existence of the function
$\bm{h}(\cdot)$, and hence allows us to obtain the
NPCE. Lemma~\ref{lem:property2} states that $\bm{h}(\cdot)$ satisfies
all the properties required for a standard interference function. The
uniqueness and iterative computation of $\bm{p}$ then follow from
Lemma~\ref{lem:yates} with the standard interference function
$\bm{h}(\cdot)$.
\end{IEEEproof}

\begin{remark}
So far we have assumed that there is no maximum power constraint imposed for any element of power $\bm{p}$. If such power constraints are imposed, then a so-called standard constrained interference function defined in \cite{Yates95} can be used instead to perform the IAP, in which the output of each iteration is set to the maximum power constraint value, if that returned from $\bm{h}$ is higher. This type of iteration converges to a unique fixed point \cite[Corollary~1]{Yates95}.
\end{remark}


\subsection{Characterization on Implementability of Load}
\label{sec:implementability}

Theorem~\ref{thm:cond_load} provides a monotonicity result for load implementability. We recall that a load vector $\loadv$ is said to be implementable if there
exists power $\bm{p}$ such that the NLCE holds.

\begin{theorem}\label{thm:cond_load}
Consider two load vectors with $\loadv'\geq \loadv$ and $\loadv'\neq \loadv$.
If $\loadv$ is implementable, then $\loadv'$ is implementable. Moreover, the respective corresponding powers $\bm{p}$ and $\bm{p}'$ satisfy $\bm{p}'<\bm{p}$. \end{theorem}
\begin{IEEEproof}
Suppose $\loadv$ is implementable, i.e., there exists power $\bm{p}$
such that the NPCE (or equivalently the NLCE) holds.  From Theorem~\ref{thm:cond_load}, $\bm{h}(\cdot)$
is a standard interference function.
We shall prove that $\loadv'$ is also implementable, i.e., $\bm{p}'$ exists.

Before we consider the general case of $\loadv'\geq \loadv$, we first
focus on the special case that strict inequality holds only for the
first element (with re-indexing if necessary), i.e.,
$\loadv=[\load_1, \load_2, \cdots, \load_n]^T$ and $\loadv'=[\load_1', \load_2, \cdots, \load_n]^T$  with
$\load_1'>\load_1$.
We now use the asynchronous IAP \eqref{eqn:IAPasy} with load $\loadv'$, and we set the initial power as $\bm{p}^0 = \bm{p}$.
Our objective is to show that the power converges to $\bm{p}'$ that satisfies the NPCE with $\bm{p}'<\bm{p}$.

Consider the asynchronous IAP \eqref{eqn:IAPasy} with outer iteration $\ell=1$ and inner iteration $i=1,2,\cdots, n$:
\bi
\item For $i=1$: 
Consider the NLCE for cell $1$ with the original load $\loadv$ and power $\bm{p}$:
\be\label{eqn:temp3}
\load_1= \sum_{j\in\mathcal{J}_1}\frac{r_{1j}}{\log\left(1+\frac{p_1 g_{1j}}{\sum_{k\geq 2} p_k g_{kj} \load_k+\sigma^2 }\right)}.
\ee
In the first iteration, $x_1$ and $p_1$ are updated by the actual load of interest $x_1'$ and the iterated power $p_1^1$, respectively, with other load and power unchanged.
Since $x'_1>x_1$,  we must have $p_1^1<p_1$.

From the proof in
Lemma~\ref{thm:fullload_is_opt}, the energy
$e_1\triangleq p_1 \load_1$ with $p_1, \load_1$ given by
\eqref{eqn:temp3} satisfies $\partial e_1/ \partial p_1 >0$. Since
$\partial e_1/ \partial x_1 = \partial e_1/ \partial p_1 \cdot
\partial p_1/ \partial x_1$ and clearly $\partial p_1/ \partial
x_1<0$, we get $\partial e_1/ \partial x_1<0$. Thus, $p_1^1 \load_1' <
p_1 \load_1$.


\item For $i=2$: We shall show that the iterated power satisfies $p_2^1< p_2^0 = p_2$. The NLCE for cell $2$ with the original load $\loadv$ and power $\bm{p}$ can be written as:
\ben
\load_2= \sum_{j\in\mathcal{J}_2}\frac{r_{2j}}{\log\left(1+\frac{p_2 g_{2j}}{p_1 g_{1j} \load_1  +  \sum_{k\geq 3} p_k g_{kj} \load_k+\sigma^2 }\right)}
\een

Upon updating cell 2, we have updated $\load_1, p_1$ to the newly iterated $\load_1', p^1_1$, respectively. Since $p_1^1 \load_1' < p_1 \load_1$ as mentioned earlier, $p^1_2<p_2$.

\item For $i\geq 3$: For subsequent iterations, it can be shown similarly that  $p_i^1< p_i^0 = p_i$ for $i=3,\cdots,n$. This completes the first outer iteration.
\ei
At this point, we get $\bm{p}_n^1 < \bm{p}$. It can be similarly shown that $\bm{p}_n^{\ell+1} < \bm{p}_n^{\ell}$ for $\ell>1$.

For large number of iterations $L$, the decreasing sequence
$\bm{p}_n^{0}, \bm{p}_n^{1}, \cdots$ must converge since
$\bm{p}_n^{\ell}\geq 0$ (i.e., it is bounded from below) for any $\ell$ due to the positivity of the standard
interference function. Thus, the power solution exists, i.e.,
$\loadv'$ is implementable.

At convergence, we have $\lim_{L\rightarrow\infty} \bm{p}_n^L =
\bm{p}' < \bm{p}$. So far we have assumed that only one element of the
load is strictly increased. In general, if more than one load element
is increased, repeating the argument sequentially for every
such element proves that power exists and is decreased.  Thus, in
general $\loadv'$ is implementable for $\loadv'\geq \loadv$, where $\bm{p}'<\bm{p}$.
\end{IEEEproof}

From Theorem~\ref{thm:cond_load}, we also obtain the equivalent result that $\loadv$ is not implementable if $\loadv'$ is not implementable.

The next theoretical characterization is on the implementable load
region $\mathcal{L}$ over all non-negative power vectors for any given
satisfiable rate $\bm{r}$, i.e., $\mathcal{L}\triangleq \{\loadv\geq
{\bm 0}: \loadv=\bm{f}(\loadv; \varspace), \bm{p}\geq \bm{0}\}$.
Theorem~\ref{thm:loadregionisopen} states that the boundary of this
region is open. The norm $\|\cdot\|$ in Theorem~\ref{thm:loadregionisopen} can be any norm, e.g., the 2-norm $\|\cdot\|_{2}$ or the maximum norm $\|\cdot\|_{\infty}$.

\begin{theorem}\label{thm:loadregionisopen}
Suppose load $\loadv$ is implementable with power $\bm{p}$ and rate
$\bm{r}$.  Then there exists $\delta>0$, such that any load
vector $\loadv'$ with $\|\loadv' - \loadv\| \leq \delta$ is implementable. Moreover, the
implementable load region $\mathcal{L}$ is open.
\end{theorem}
\begin{IEEEproof}
Let $\widetilde{\bm{p}}=\beta\bm{p}$ with $\beta>1$, and let the
corresponding load satisfying the NLCE  with
rate $\bm{r}$ be $\widetilde{\loadv}$.
Note that $\widetilde{\loadv}$ exists, because the existence of
load does not depend on power (cf. Lemma \ref{lem:uniqueness}).
By applying the IAL in
\re{eqn:algo} to obtain $\widetilde{\loadv}$ (using power
$\widetilde{\bm{p}}$) with the initial load set as $\loadv^0=\loadv$,
it can be easily checked that the load vector decreases
in every iteration. Since $\widetilde{\loadv}>\bm{0}$, the
iterations must converge to $\widetilde{\loadv} = \lim_{L\rightarrow
\infty} \loadv^L < \loadv$.  By Theorem~\ref{thm:cond_load}, any
$\loadv' \geq \widetilde{\loadv}$ is implementable.
As $\widetilde{\loadv} < \loadv$, there is an implementable neighbourhood of
$\loadv$. That is, there exists $\delta >0$, for
which any load vector $\loadv'$ satisfying $\|\loadv' -\loadv\| \leq \delta$
is implementable.  Since the result holds for any $\loadv$ in
$\mathcal{L}$, it follows that $\mathcal{L}$ is open.
\end{IEEEproof}

\subsection{Algorithm for Optimal Power Vector}
\label{sec:optpowervect}

By Theorem~\ref{thm:findoptp}, we can use the IAP to compute the
optimal power $\bm{p}^{\star}$ for \eqref{eqn:optpow} in
Theorem~\ref{thm:optsol} for any given implementable load $\loadv^{\star}$.
We recall that to minimize the energy we set $\loadv^{\star}=\bm{1}$ (Theorem~\ref{thm:optsol}).
To obtain the output of the function
$\bm{h}(\cdot)$ in each step of the IAP, bisection search is able to
determine the power $p_i$ such that $\eta_i(p_i)=1$ (see
Remark~\ref{rem:computeh}). Putting together the theoretical insights
results in the following formal algorithmic description (Algorithm
\ref{alg:power}) for computing $\bm{p}^{\star}$.

\begin{algorithm}[ht!]
{\bf Given}: \\
- target load vector 
${\loadv^{\star}}=[\load^{\star}_1,\load^{\star}_2,\cdots, \load^{\star}_n]^T$ \\
- rate vector $\bm{r}$ such that $\rho(\bm{\Lambda}(\bm{r})) < 1$ \\
- arbitrary initial power vector ${\bm p}$ \\
- tolerance $\epsilon>0$ 
\\
{\bf Output}: ${\bm p}^{\star}$ with $\loadv^{\star} = \bm{f}(\loadv^{\star}; \bm{r}, {\bm p}^{\star})$

\begin{algorithmic}[1]
\STATE Initialize  $\loadv\leftarrow \bm{f}(\bm{\loadv^{\star}}; \bm{r}, {\bm p}) $.
\WHILE { $\Vert{\bm x}-\loadv^{\star}\Vert_{\infty} > \epsilon$} \label{line:loop}

\FOR{$i=1:n$} \label{line:cell}

    \STATE $p^{\sf left}_i \gets \xi$ for any $\xi$ such that $\eta_i(\xi)>1$
    \STATE $p^{\sf right}_i \gets \psi$ for any $\psi$ such that $\eta_i(\psi)<1$ 

\label{line:bistart}

    \WHILE {$|\eta_i(p_{i})  - 1|  > \epsilon$}
         \IF {$ \eta_i(p_{i})  \leq  1$ } \label{line:bisectstart}
              \STATE $p^{\sf right}_i \leftarrow p_i$       \label{line:update1}
         \ELSIF {$ \eta_i(p_{i})  > 1$}
              \STATE $p^{\sf left}_i \leftarrow p_i$  \label{line:update2}
         \ENDIF
\STATE  {\bf end if}
    \STATE $ p_i \leftarrow (p^{\sf left}_i+p^{\sf right}_i)/2$ 
    \ENDWHILE
        \label{line:bisectend}
    \STATE  {\bf end while}   \label{line:biend}
\ENDFOR
\STATE  {\bf end for}
\STATE $\loadv\leftarrow \bm{f}(\bm{\loadv^{\star}}; \bm{r}, {\bm p}) $ \label{line:xupdate}
\ENDWHILE
\STATE  {\bf end while}
\STATE  $ {\bm p}^{\star} \gets {\bm p}$, return ${\bm p}^{\star}$

\end{algorithmic}
\caption{IAP algorithm for computing optimal power.  \label{alg:power}}
\end{algorithm}
}

Algorithm \ref{alg:power} solves the NPCE for given $\bm{x}^{\star}$,
by iteratively updating the power vector and re-evaluating the resulting
load $\bm{f}(\bm{\loadv}^{\star}; \bm{r}, {\bm p})$.  The bulk of the
algorithm starts at Line~\ref{line:loop}.  The outer loop terminates
if the load vector $\bm{x}$ has converged to $\bm{x}^{\star}$.  For
each outer iteration, the inner loop is run starting at
Line~\ref{line:cell}, for which  the power vector for each cell $i$ is updated.
In each update, the power range is first initialized to $[\xi, \psi]$, where $\xi<\psi$, such that $\eta_i(\xi)>1$ and $\eta_i(\psi)<1$.  Since the function $\eta_i(\cdot)$ is a strictly
decreasing function, the bisection search from
Lines~\ref{line:bisectstart}-\ref{line:bisectend} ensures convergence
to the unique solution for $\eta_i(p_i)=1$, or equivalently, the value of
$h_i(\bm{\bar p}_i; \loadv, \bm{r})$. Load re-evaluation is then carried out
in Line \ref{line:xupdate}.

\section{Numerical Evaluation}\label{sec:numerical}
\subsection{Simulation Setup}

In this section, we provide numerical results to illustrate the
theoretical findings.  The simulations have been performed for a
real-life based cellular network scenario, with publicly available
data provided by the European MOMENTUM project \cite{momentum}.
The channel-gain data are derived from a path-loss model and calibrated with real measurements of signal strength in the network of a sub-area of Alexanderplatz in the city of Berlin.
The path-loss model takes into account the terrain and environment, pre-optimized antenna configuration (height, azimuth, mechanical tilt, electrical tilt); fast fading is not part of the data made available. Further details are available in \cite{momentum}.

\begin{figure}
\centering
\includegraphics[scale=\MyFigSizetwo]{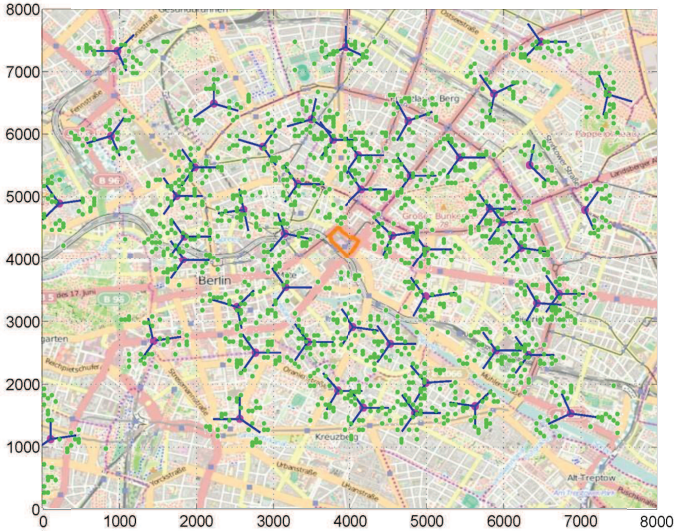}
\caption{Network layout and user distribution in an area of Alexanderplatz, Berlin. The units of the axes are in meters.  Digital Map: $\copyright$ OpenStreetMap contributors, the map data is available under the Open Database License.}
\label{berlin}
\end{figure}

The scenario is illustrated in Fig. \ref{berlin}. The scenario has 50 base station sites, sectorized into 148 cells.
In Fig. \ref{berlin}, the red dots indicate base station sites and the
green dots represent the location of users.  Most of the sites have
three sectors (cells) equipped with directional antennas. The blue
short lines represent the antenna directions of the cells.  The entire
service area of the Berlin network scenario is divided into
22500~pixels as shown in Fig. \ref{berlin}. That is, each pixel represents a small
square area, with resolution 50~$\times$~50~m$^2$, for which signal
propagation is considered uniform.  Users located in the same pixel
are assumed to have the same channel gains. In our simulations, each cell
serves up to ten randomly distributed users in its serving area as defined in the MOMENTUM data set.
The total bandwidth of each cell is 4.5~MHz. Following the LTE
standards, we use one resource block to represent a resource unit
with 180~kHz bandwidth each in the simulation.  Network and simulation
parameters are summarized in Table~\ref{tab:1}.

\begin{table}
\caption {Network and simulation parameters}\label{tab:1}
\label{parameter}
\centering
\begin{tabular}[t]{l l}
\hline
\bf{Parameter}  &   \bf{Value}   \\

Service area size & 7500 $\times$ 7500 m$^2$ \\
Pixel resolution  &  50 $\times$ 50 m$^2$  \\
Number of sites  & 50 \\
Number of cells  & 148   \\
Number of pixels  & 22500  \\
Number of users  & 1480  \\
Thermal noise spectral density  & -145.1 dBm/Hz   \\
Total bandwidth per cell & 4.5 MHz \\
Bandwidth per resource unit & 180 kHz \\
Tolerance $\epsilon$ in IAP  &  $10^{-5}$ \\
Initial power vector $\bm p$ in IAP &  $\bm1$ W  \\
\hline
\end{tabular}
\end{table}

%
%

\subsection{Results}



Our objective is to numerically illustrate the relationship among the load, power, and sum
transmission energy.  First, we consider the use of uniform load with
$\loadv=\phi \bm{1}$ for various $0<\phi\leq 1$, with $\phi=1$ being the
case of full load.  Given the load vector $\loadv$, the optimal power
solution $\bm{p}$ is then obtained by using the IAP described by
Algorithm~\ref{alg:power}.  Next, for benchmarking, we consider the
conventional scheme that employs uniform power allocation
$\bm{p}=\beta\bm{1}, \beta>0$. We choose $\beta$ that results in the
minimum sum energy subject to the constraint that the corresponding
load satisfies $\bm{0} \leq \loadv \leq \bm{1}$, as follows.  From the proof of
Lemma~\ref{thm:fullload_is_opt}, the energy (given by the product of
load and power) for each cell strictly decreases as the power strictly
decreases.  Thus, to minimize the sum energy, we choose the smallest
$\beta$ such that $\loadv \leq \bm{1}$; this can be obtained by a
bisection search starting with sufficiently small and large values of
$\beta$. For any $\beta$ under consideration, the IAL is used to
obtain the load corresponding to the power $\bm{p}=\beta\bm{1}$.


\begin{figure}
\centering
\includegraphics[scale=\MyFigSize]{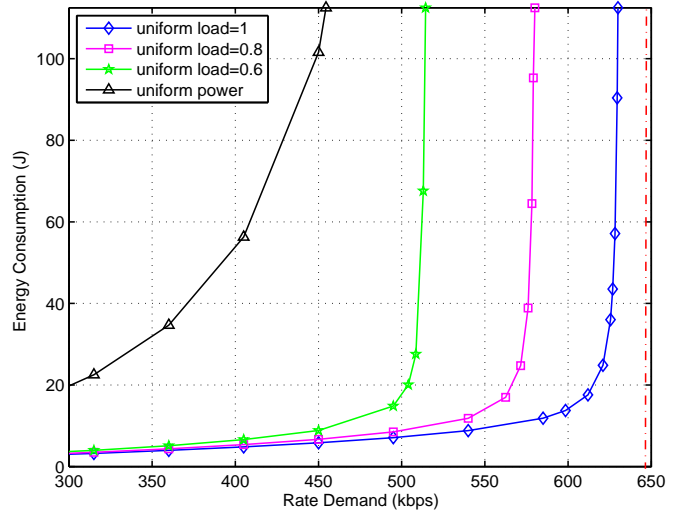}
\caption{Sum transmission energy with respect to user's rate demand.}
\label{demand}
\end{figure}

In the first numerical experiment, we consider the sum energy for
rate demand $\bm{r}= \xi \bm{1}$ with $\xi$ being successively increased,
while keeping $\bm{r}$ satisfiable.  Fig.~\ref{demand} compares the
sum energy for various uniform load levels, including full load, and
that obtained by uniform power allocation.  From Fig.~\ref{demand}, the
sum energy for all cases appears to grow exponentially fast as the
rate demand increases, approaching infinity as the rate demand increases. The vertical dotted line in Fig.~\ref{demand} corresponds to the
boundary when the rate demand is not satisfiable, i.e.,
$\rho(\bm{\Lambda}(\bm{r})) = 1$, and hence represents the upper bound for which the system can support. This behaviour is consistent with
Lemma~\ref{lem:uniqueness}.  Deploying full load achieves the smallest
sum energy, in accordance with Lemma~\ref{thm:fullload_is_opt}. The
reduction in sum energy is particularly evident in comparison to the
scheme of uniform power -- the relative saving is $90\%$ or higher for the rate demand shown in Fig.~\ref{demand}.
Conversely, for a fixed amount of sum energy, deploying full load
and optimizing the corresponding power allows for maximizing the rate demand that can be served.

%

\begin{figure}
\centering
\includegraphics[scale=\MyFigSize]{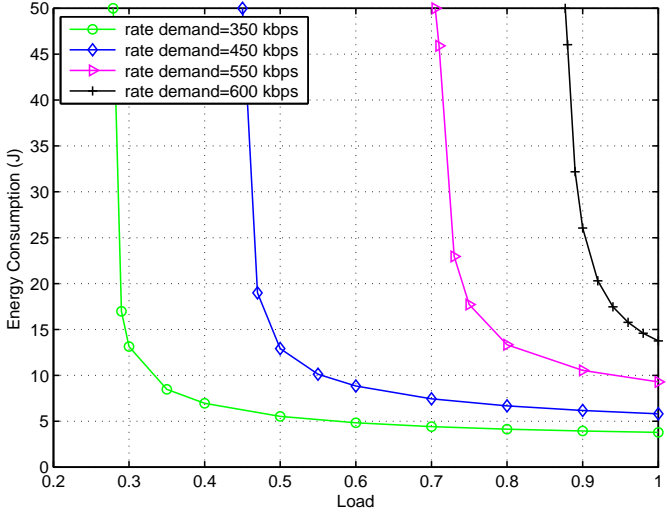}
\caption{Sum transmission energy with respect to cell load.}
\label{load}
\end{figure}

Next, we examine the energy consumption by progressively increasing
the uniform load for four rate demand levels $\bm{r}= \xi \bm{1}$ with
$\xi$ taking the values of 350~kbps, 450~kbps, 550~kbps and 600~kbps.  The results are shown in Fig.~\ref{load}.  We observe that the sum energy decreases
monotonically by increasing the load.  The reduction of sum energy
appears to be exponentially fast in the low-load regime, but is much
slower in the high-load regime.  In addition, the numerical results
reinforce the fact that some load vectors are not
implementable. In particular, it is not always possible to obtain a
power vector $\bm{p}$ for a load vector $\loadv=\phi \bm{1}$ with very
small $\phi>0$.  From Fig.~\ref{load}, the sum energy surges to
infinity when the load approaches some fixed (small) value, which
suggests that, for any $\bm{r}>0$, the load cannot become arbitrarily small,
irrespective of power.

\begin{figure}
\hspace{-3mm}
\centering
\includegraphics[scale=\MyFigSize]{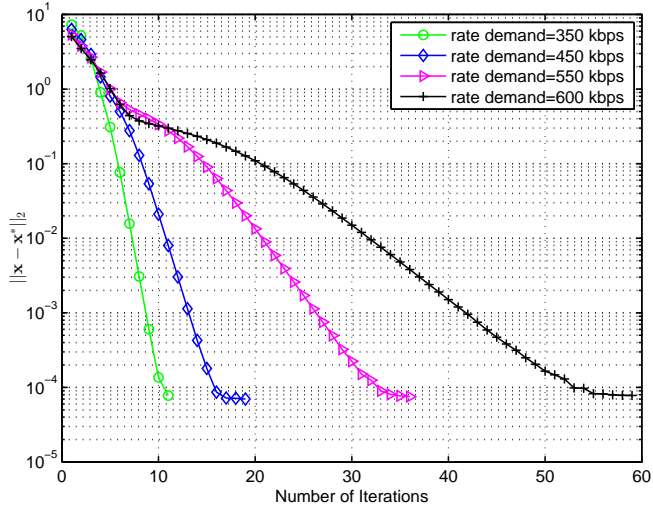}
\caption{The evolution of the Euclidean distance between the iterate $\bm{x}$ and the target $\bm{x^{\star}}$, given by the 2-norm $\|\bm{x}-\bm{x^{\star}}\|_2$, over iterations with $\bm{x^{\star}}=\bm{1}$.}
\label{converge}
\end{figure}

Furthermore, we numerically investigate the convergence
behavior of the IAP. The theoretical analysis on the convergence speed of the IAP depends on whether $\bm{h}(\cdot)$ further satisfies some property such as contractivity \cite{FeyzmahdavianTWCOM12}, for which linear rate of convergence can be shown to hold.
In Fig.~\ref{converge}, we set the target load
vector $\bm x^{\star}=\bm 1$ and the initial power vector $\bm p=\bm1$ Watt, with rate demand $\bm{r}= \xi \bm{1}$ where
$\xi \in \{350, 450, 550, 600\}$ kbps. The Euclidean distance between the iterate
$\bm{x}$ and target $\bm{x}^\star$ is given by the 2-norm
$\|\bm{x}-\bm{x^{\star}}\|_2$. The
evolution of $\|\bm{x}-\bm{x^{\star}}\|_2$ for the four different rate demand cases
is illustrated in Fig.~\ref{converge}.  We consider the algorithm converged if the largest error between the load iterate and the target load is less than $\epsilon=10^{-5}$, i.e., if $\|\loadv - \loadv^{\star}\|_{\infty} \leq \epsilon$.
For the four rate demand cases, convergence is reached after 11, 19, 36 and
59 iterations, respectively. Given the size of the network (148 cells),
the values are moderate. Also, we notice that
when the rate demand increases, more iterations are required for
convergence with a longer tail-off.
This is mainly because a high rate demand means that, in general, the NPCE
is operating in the high SINR regime. The amount of progress in load
in an IAP iteration is mainly dependent on the denominator in
\eqref{eqn:nonlinearprob}.
For high SINR regime, the relative change in load
is lesser due to the logarithm
operator, thus slowing down the progress.
Moreover, the number of iterations depends on the initial power point.
In general, fewer iterations are required if the starting power point is closer to the optimum.
Note that no matter what the rate of convergence is, the convergence of the IAP is guaranteed by Theorem~\ref{thm:findoptp}.
The convergence speed depends also on other factors, e.g., the scale of the network (in terms of the number of BS and users), the rate demand and the choice of $\epsilon$.
An explicit characterization of this dependence is beyond the scope of this paper.

In case of the presence of some time constraints in a practical
application, the IAP may be terminated before full convergence is
reached. Thus, the capability of delivering a load-feasible and
close-to-convergent solution within few iterations is of
significance. It can be seen in Fig.~\ref{converge} that a majority of
the iterations is due to the tailing-off effect -- the load vector is
in fact close to the target value within about half of the
iterations. For all the rate demand levels, convergence is in effect
achieved in less than 20 iterations; this is promising for the
practical relevance of the proposed IAP scheme.
Finally, to ensure that the load is strictly less than full load for practical implementation, we may set $\bm{x}^{\star}=(1-\epsilon')\bm{1}$ with $\epsilon' > \epsilon$.

\begin{figure}
\centering
\includegraphics[scale=\MyFigSize]{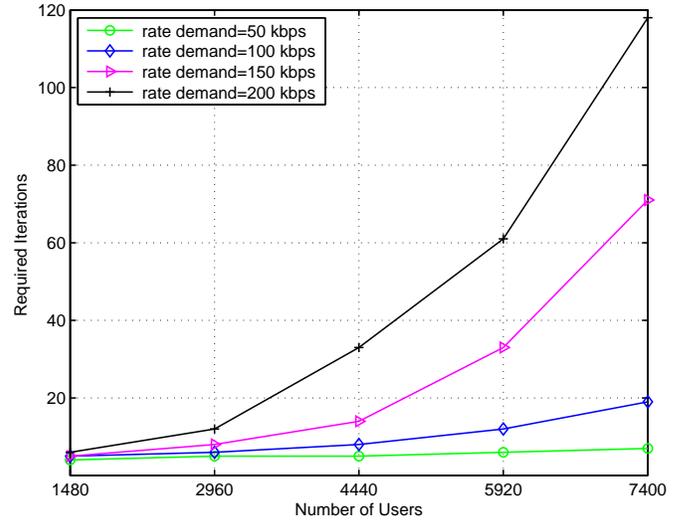}
\caption{The number of iterations required to achieve convergence for different number of users.}
\label{user_iteration}
\end{figure}

Fig.~\ref{user_iteration}  illustrates the the number of iterations required for the load to converge, i.e., until $\|\loadv - \loadv^{\star}\|_{\infty} \leq \epsilon$, with $\epsilon=10^{-5}$. The total number of users is increased from the current $1480$ (corresponding to $10$ users per cell) to $7400$ (corresponding to $50$ users per cell).
We have set the rate demand as $\bm{r}=\xi \bm{1}$ with $\xi\in \{50, 100, 150, 200\}$~kbps, because the original case of $\xi\in\{350, 450, 550, 600\}$ is no longer satisfiable for $7400$ users.
From Fig.~\ref{user_iteration} , we see that the number of iterations increases as the number of users increases, and the rate of increase is higher if the number of users is large or the demand is high. Hence, for systems that support high data rate or large number of users, more computational resources are needed to implement the algorithm.


\section{Conclusion}\label{sec:con}
We have
obtained some fundamental properties for the cellular network modeled
by a non-linear load coupling equation (NLCE), from the perspective of
minimizing the energy consumption of all the base stations. To obtain
analytical results on the optimality of full load, and the computation
and existence of the power allocation, we have investigated a dual to
the NLCE, given by a non-linear power coupling equation
(NPCE). Interestingly, although the NPCE cannot be stated in
closed-form, we have obtained useful properties that are instrumental
in proving the analytical results.
Our analytical results suggest that in load-coupled OFDMA networks or more specifically LTE networks, the maximal use of bandwidth and time resources over power leads to the highest energy efficiency.
In the literature, the maximal use of resources is typically suggested to maximize the network throughput; our work gives a similar conclusion but from a different and complementary approach of minimizing energy.
To implement the solution, the load and power solutions have to be computed and sent to all base stations for implementation. Hence, some level of coordination has to be set up in practice. In this paper, we have assumed the use of ideal power amplifier and that the users' associations to the base stations are given. The effects of non-linear power amplifier and the problem of user association may be considered as future work.

\section*{Acknowledgements}
We would like to thank the anonymous reviewers for their valuable comments and suggestions. The work of the second author has been supported by the
Link\"{o}ping-Lund Excellence Center in Information Technology (ELLIIT), Sweden. The work of the third author has been supported by the Chinese Scholarship Council (CSC) and the overseas PhD research internship scheme from Institute for Infocomm Research (I$^2$R), A*STAR, Singapore.


\appendix 

\begin{lemma}[Theorem~2, \cite{HoYuanSun13}]\label{thm:d_inc_with_x}
Consider the NLCE \eqref{eqn:nonlinearprob} with power $\bm{p}$ fixed.
Given the rate vectors $\bm{r}'$ and $\bm{r}$ with $\bm{r}' \geq \bm{r}$ and $\bm{r}' \neq \bm{r}$, the corresponding load vectors  $\loadv'$ and $\loadv$ satisfy $\loadv'>\loadv$.
\end{lemma}

We omit the proof of Lemma~\ref{thm:d_inc_with_x}, which is given in \cite{HoYuanSun13}.

\begin{lemma}\label{thm:constraintequality}
For Problem $P0$, the optimal rate vector satisfies $\bm{r}^{\star} = \bm{d}_{\mathrm{min}}$.
\end{lemma}
\begin{IEEEproof}
Suppose that at optimality, there exists at least one rate element $r^{\star}_{ij}$ that is strictly greater than its corresponding (minimum) rate demand $d_{ij,\mathrm{min}}$.
Taking the power to be fixed as $\bm{p}^{\star}$, if we decrease $r^{\star}_{ij}$ to $d_{ij,\mathrm{min}}$, then the load will strictly decrease while satisfying the constraint \eqref{eqn:P0c1} by Lemma \ref{thm:d_inc_with_x}. Thus, the objective function value decreases. This contradicts the optimality of $r^{\star}_{ij}$. Thus $\bm{r}^{\star} = \bm{d}_{\mathrm{min}}$.
\end{IEEEproof}



\end{document}